# Laser emission with excitonic gain in a ZnO planar microcavity

T. Guillet[1,2], C. Brimont[1,2], P. Valvin[1,2], B. Gil[1,2], T. Bretagnon[1,2], F. Médard[3], M. Mihailovic[3], J. Zúñiga-Pérez[4], M. Leroux[4], F. Semond[4], S. Bouchoule[5]

*1 : Université Montpellier 2, Laboratoire Charles Coulomb UMR 5221, F-34095, Montpellier, France*

*2 : CNRS, Laboratoire Charles Coulomb UMR 5221, F-34095, Montpellier, France*

*3 : LASMEA, UMR 6602, UBP – CNRS, 24 Avenue des Landais, F-63177 Aubière, France*

*4 : CRHEA – CNRS, Rue Bernard Grégory, F-06560 Valbonne, France.*

*5 : LPN – CNRS, Route de Nozay, F-91460 Marcoussis, France.*

**Abstract**

The lasing operation of a ZnO planar microcavity under optical pumping is demonstrated from T=80 K to 300 K. At the laser threshold, the cavity switches from the strong coupling to the weak coupling regime. A gain-related transition, which appears while still observing polariton branches and, thus, with stable excitons, is observed below 240K. This shows that exciton scattering processes, typical of II-VI semiconductors, are involved in the gain process.

Blue and ultraviolet operating lasers are attracting a strong interest due to the wide range of foreseen applications (display, optical data storage, biomedical). They are based on GaN and ZnO semiconducting materials and various kinds of optical resonators are presently investigated : Fabry-Perot modes in planar microcavities and in nanowires [1, 2], whispering gallery modes in microwires [3] and microdisks [4], cavities in photonic crystal membranes [5], as well as random lasing in ZnO powders [6] and waveguides [7]. Three main mechanisms are invoked as responsible for gain : (i) an electron-hole plasma is involved in recently demonstrated GaN and ZnO vertical cavity surface emitting lasers (VCSELs) [8, 9] and ZnO nanowire lasers [10]; (ii) exciton scattering processes are claimed to mediate gain in ZnO nanowires [2]; (iii) when excitons and photons are in the strong coupling regime up to high densities, polariton lasing is observed, as demonstrated in GaN planar microcavities [11]. These different alternatives show that a deeper understanding of the gain operation mechanism is still required, especially in ZnO microcavities in which the frontier between each of these regimes, if present, is not yet clearly established. Furthermore, the exciton-mediated processes, which are specific to II-VI materials and have been thoroughly studied 40 years ago in bulk samples (see [12] for a critical review), are still the object of controversies in microlasers.

In the present work, we demonstrate lasing operation from 80 K to 300 K in a ZnO planar microcavity that exhibits a strong exciton-photon coupling at low exciton density and that switches into the weak coupling regime at lasing threshold. The gain mechanism is proved to be exciton-mediated at threshold, as shown by the coexistence of strong exciton-photon coupling and a gain-related transition when the cavity mode is brought to a large negative detuning.

The microcavity consists in a 5λ/4 ZnO active layer embedded between a bottom (Ga,Al)N/AlN DBR and a top (Si,O)/(Si,N) DBR. The nitride DBR and the ZnO layer are grown by MBE on a Si(111) substrate, whereas the dielectric DBR is realized by PECVD [13, 14]. The number of pairs of each mirror is respectively 13 and 12, leading to a nominal quality factor of about 500 for the cavity mode. Experimentally, the quality factor of the cavity depends on the size of the excitation spot : it amounts 300 for ~100 µm spots, and reaches 450 for ~2 µm spots. It is therefore close to the nominal value but only locally, as inhomogeneous broadening due to in-plane photonic disorder on length scales smaller than 100 µm tends to degrade it [15].

In the linear regime, the microcavity exhibits the regime of strong exciton-photon coupling both at low and room temperature. The study of the angle-resolved reflectivity of the half-cavity, i.e. without the top dielectric DBR, revealed a large Rabi splitting of 130meV at 300K [14]. A smaller value (105±10 meV) is deduced from the analysis of the angle-resolved photoluminescence of the actual full cavity, the diminution of the Rabi splitting being due to the larger penetration length in the dielectric DBR than in air.

In this paper, the emission of the microcavity has been studied under strong excitation density. The cavity is excited by an OPO (Opotek Vibrant UV-HE-LD, pulse duration: 3 ns; repetition rate 10 Hz, spot diameter 50 µm) at 3.78eV, and an incidence angle of 30°, corresponding to the first Bragg mode on the high energy side of the stop-band. Figure 1 presents a series of PL spectra with an increasing excitation power density at T=300K. At a low excitation density, the spectrum consists in a sharp peak at 3.20 eV related to the lower polariton branch. As the excitation density is increased above a threshold value of 0.8 $J.cm^{-2}$, a second sharp peak that grows non-linearly appears, (see Figure 1(b)). Its energy corresponds to the value of the bare photon mode in the absence of strong coupling with excitons. This peak is attributed to lasing, i.e. to the VCSEL operation of the microcavity. The LPB transition remains visible after threshold, and is probably related to polariton emission at the edge of the excitation spot, where the photo-created density is smaller than in the center, as already mentioned in the case of a strongly coupled GaAs microcavity switching to VCSEL operation [16].

The temperature dependence of the observed spectra around threshold provides valuable information on the operation mechanism of the laser. Figure 2(a) presents the power dependence of the PL spectrum at T=80 K. At low excitation density, the spectrum now consists in one sharp peak at 3.22 eV related to the lower polariton branch, a weaker and broader transition at 3.364 eV at the energy of free excitons, and an additional sharp peak at 3.342 eV. This last feature is also present under continuous wave (CW) excitation and presents a slight angular dispersion as shown in Figure 2(d). It is a higher order mode of the cavity ($7\lambda/4$ instead of $5\lambda/4$) strongly coupled to excitons, as predicted in a previous theoretical work on ZnO microcavities [17]. It appears thanks to the strong variation of the refractive index nearby the excitonic transitions in ZnO [18-20] i.e. the large excitonic oscillator strength. A similar effect has been observed in ZnO nanowire lasers, where the free spectral range strongly decreases nearby the exciton energy [1, 2]. As the temperature is increased the exciton homogeneous broadening attenuates the variation of the refractive index, resulting in a $7\lambda/4$ mode that is broader but still visible at 240K (Figure 2(b)), and disappears at 300 K (Figure 1). The figure 2(d) also shows the dispersion of the uncoupled photon modes, and of the polaritons, determined from a coupled oscillator model. The obtained energy of the cavity mode is $3.243\pm0.01$ eV.

At 80K, when the excitation power is increased, an additional transition (G) appears in the PL spectrum (see Figure 2(a)), shifting from 3.305 to 3.28 eV. Its temperature dependence is plotted in Figure 3 and compares well with previously reported stimulated emission features in ZnO, which are usually attributed in bulk ZnO to exciton-exciton scattering processes at low temperature and low density, and gain in the electron-hole plasma (EHP) at higher density (see [12] for an exhaustive review). This peak does not disperse with angle or with detuning, as shown in Figure 2(e) and is, therefore, not related to the cavity. It could be due either to amplified spontaneous emission (ASE) or to random lasing in the plane of the cavity, as recently reported for ZnO waveguides [7], which would scatter to the LPB and Bragg modes due to interface roughness and cracks. When the excitation density is further increased, the lasing mode (L) is observed at the calculated energy of the cavity mode (Figure 2(e)).

The energies of LPB and 7λ/4 modes at low density and of the lasing mode at threshold are also reported in Figure 3 for an initial detuning at low temperature of -120 meV. The lasing mode only weakly shifts with temperature, as expected for a photon mode only sensitive to the refractive index. The LPB and 7λ/4 modes partly follow the Varshni law for ZnO excitons, which supports the fact that they are in the strong coupling regime.

Let us now discuss the role of the excitons and the electron-hole plasma in the gain mechanism. Figure 2(c) presents the power dependence of the PL spectrum for a larger negative detuning compared to figures 2(a) and 2(b) and the same range of excitation. No lasing is observed in the same excitation range although the stimulated emission feature is as strong as for smaller detunings, albeit too far from the cavity mode to observe lasing. More interestingly, the LPB and 7λ/4 modes are kept sharp over the whole density range, which proves that the strong coupling regime of excitons with photons is maintained at the highest carrier density when lasing is not observed. This implies that, within the investigated density range, the electron-hole system is kept below the Mott transition to an electron-hole plasma, and that the gain at 3.3 eV can result from P-band exciton scattering processes [21, 22]. On the contrary, when lasing is observed (Fig. 2(a) and 2(b)), the LPB and 7λ/4 modes rapidly broaden above threshold and the system transits into the weak coupling regime. It has to be noticed that it is difficult to directly estimate and compare carrier densities, since they depend on the carrier lifetime in the ASE or random laser regimes and this is unknown. The energy shift of the LPB mode is comparable in both cases (+4 meV), showing that the densities are similar. Moreover, if we consider the energy shift of the peak G as another indicator of the carrier density [21], the difference between densities in figures 2(a) and 2(c) is of the order of 50%. We can therefore conclude that at T=80K in our ZnO microcavity, gain is not initially due to a degenerate EHP, which is the standard gain mechanism in III-V VCSELs. Furthermore, this behaviour and its associated mechanism continue to operate similarly up to T=240 K. Unfortunately, from our current data we cannot distinguish the gain mechanism at T=300 K.

As a conclusion, we have demonstrated lasing in a ZnO microcavity operating from 80 K to room temperature. Below T=240 K, the gain mechanism implies initially exciton-scattering processes, as proved by the coexistence of the strong coupling regime and a gain-related transition at negative detuning. This mechanism is different from the standard VCSEL operation, which only relies on gain in a degenerate EHP. This supports the recent interpretation of gain mechanism in ZnO nanowires [2] in terms of exciton scattering processes.

This work was funded by the ANR project “ZOOM” (ANR-06-BLAN-0135) and the European project Clermont4 (FP7-PEOPLE-ITN-2008 235114).

FIGURES

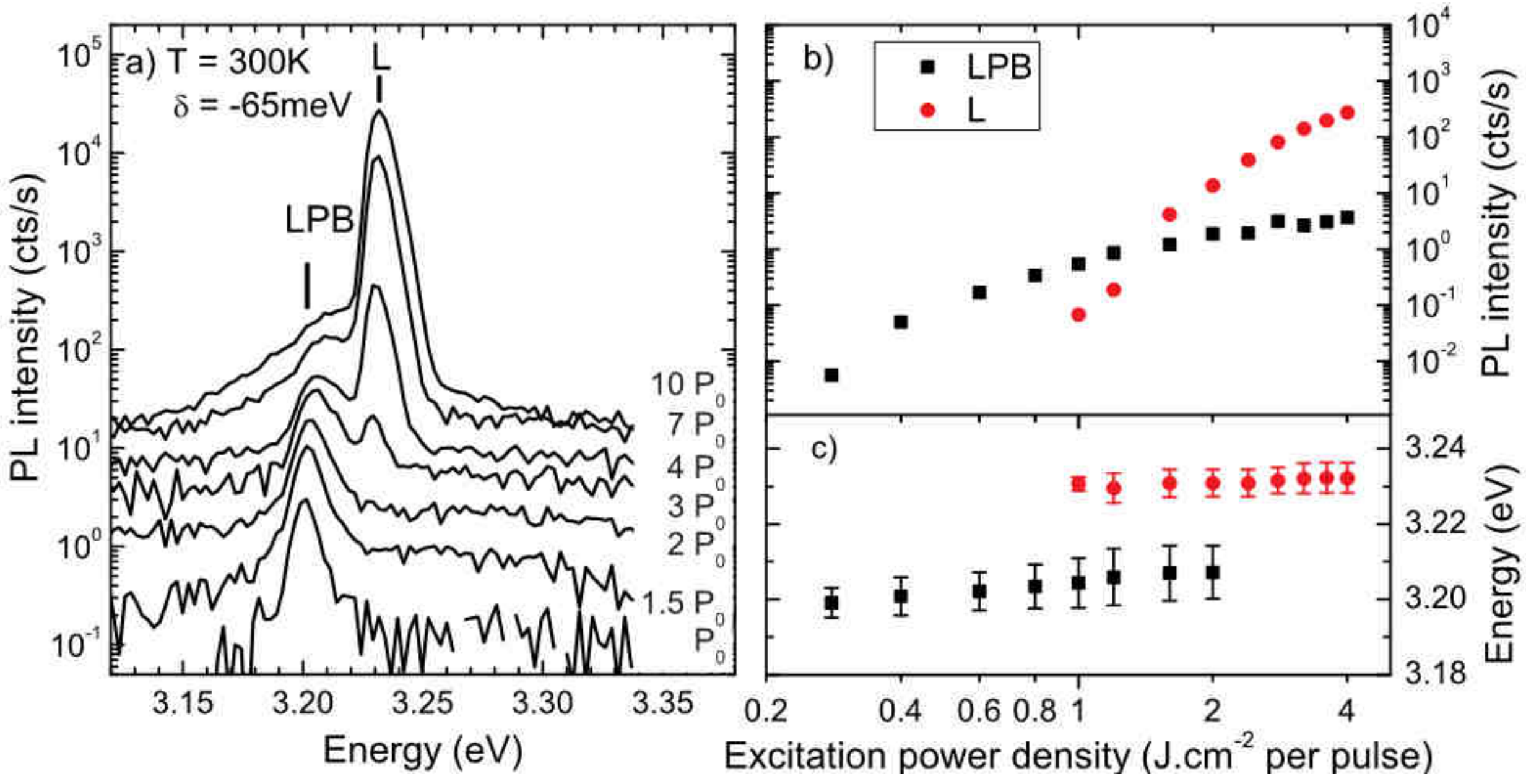

Figure 1 : (a) Photoluminescence at T=300 K and a detection angle of 0°, as a function of the excitation power density ($P_0$=0.4 J.cm$^{-2}$ per pulse); the lower polariton branch (LPB) and the lasing mode (L) are indicated; (b,c) Amplitude and energy of the two observed transitions.

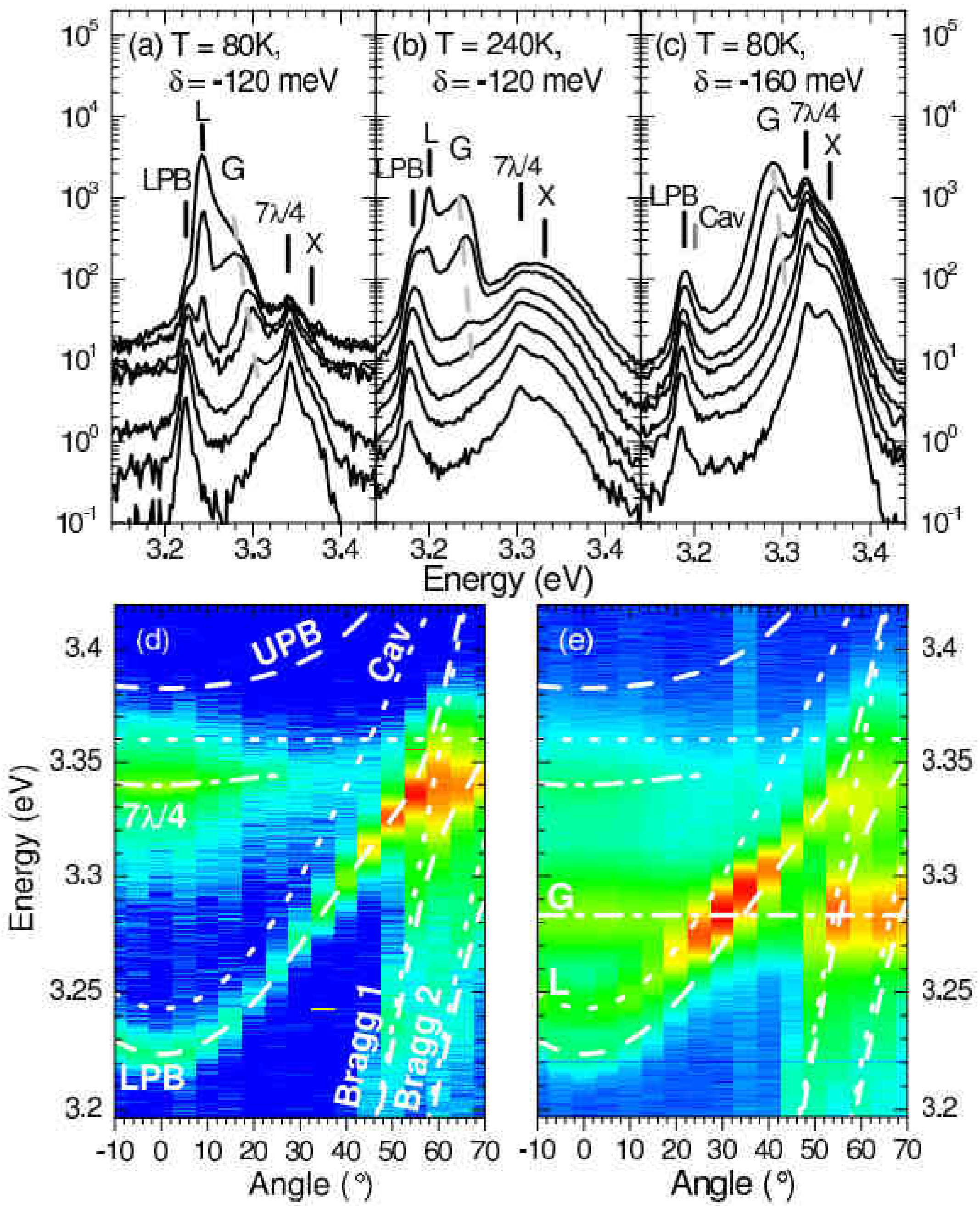

Figure 2 : (a-c) Photoluminescence recorded at 0° as a function of the excitation power density for various temperature and detuning conditions; the excitation power density increases from $P_0$ = 0.4 J.cm$^{-2}$ to 10 $P_0$, similarly to Figure 1. The lower polariton branch (LPB), lasing mode (L), gain feature (G), 7λ/4 polariton mode (7λ/4) and exciton energy (X) are indicated; the expected uncoupled cavity mode is also shown for fig. 2(c). (d,e) Angle-resolved photoluminescence spectra below (d, P = $P_0$) and above (e, P = 5 $P_0$)

the lasing threshold, in the same conditions as in Figure 2.a; the PL intensity is represented in a logarithmic color scale. The dispersion of uncoupled photon and exciton modes are indicated as dotted lines; the dashed lines represent the polariton modes obtained from a coupled oscillator model; the gain feature appears as dashed-dotted line.

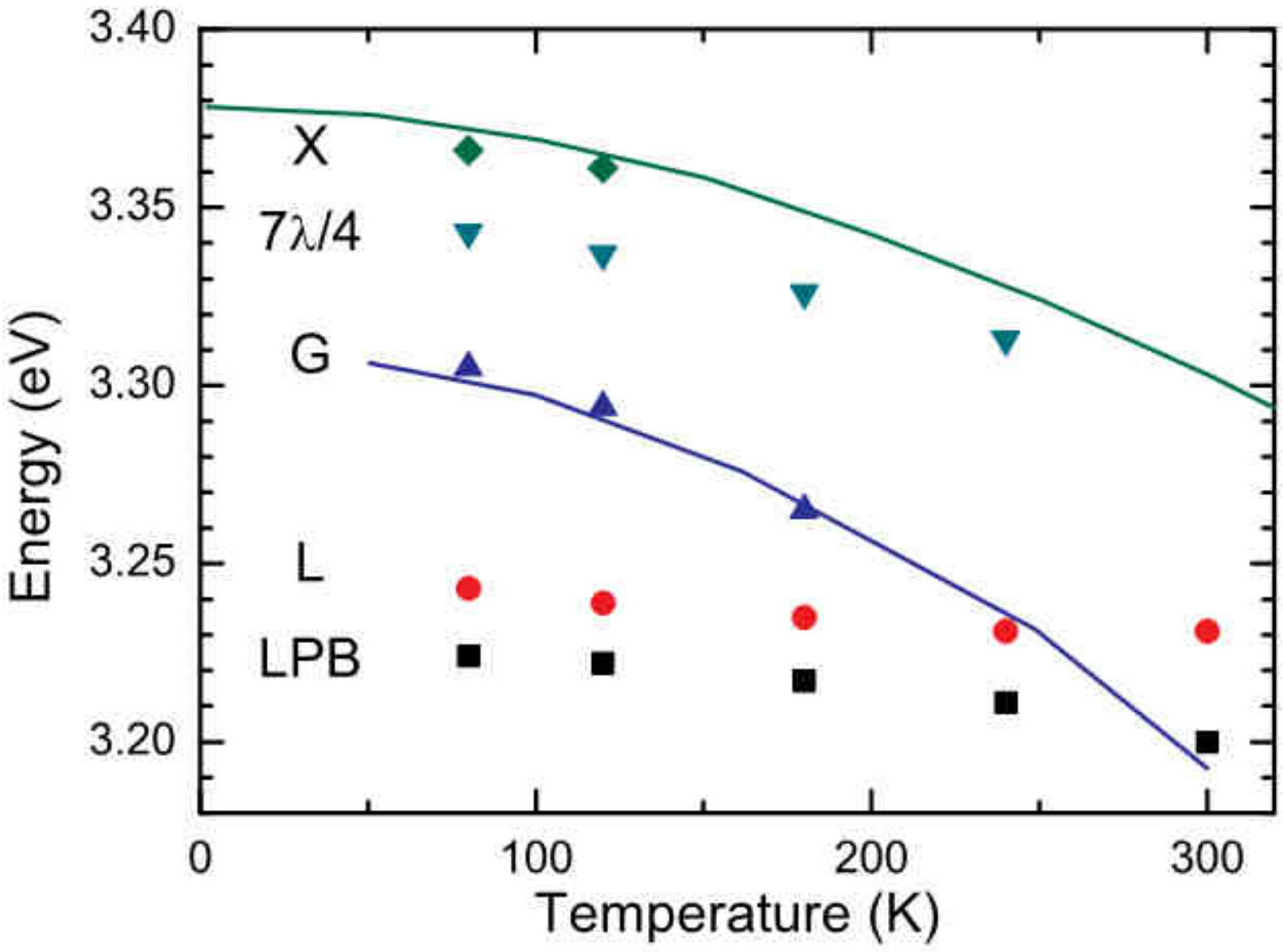

Figure 3 : Energies (symbols) of the observed transitions as a function of temperature : lower polariton branch (LPB), lasing mode (L), gain feature (G), 7λ/4 polariton mode (7λ/4), and exciton (X); the Varshni law for zinc oxide excitons and the reported values for the gain feature [12] are shown as green and blue lines respectively.